\documentclass[3p,twocolumn,breaklinks,hyphens]{elsarticle}

\usepackage{natbib}
\usepackage{amsmath,amssymb,amsfonts} 
\usepackage{algorithmic} 
\usepackage{graphicx} 
\usepackage{textcomp} 
\usepackage{hyperref} 
\usepackage{float} 
\usepackage{verbatim} 
\usepackage{tabularx}

\usepackage{pifont} 
\usepackage{ragged2e}

\usepackage{url} 
 
\usepackage{breakurl}

\usepackage{hyperref}

\usepackage{colortbl} 
\usepackage[normalem]{ulem}

\newcommand{\xmark}{\ding{55}} 
\def\BibTeX{{\rm B\kern-.05em{\sc i\kern-.025em b}\kern-.08em T\kern-.1667em\lower.7ex\hbox{E}\kern-.125emX}} 
\definecolor{orange}{rgb}{1,0.647,0}

\begin{document}

\title{Vulnerability of Blockchain Technologies to Quantum Attacks}

\author[1]{Joseph J. Kearney } \ead{jjk30@kent.ac.uk} 
\author[1]{Carlos A. Perez-Delgado \corref{cor1} } \ead{c.perez@kent.ac.uk} 

\address[1]{School of Computing, University of Kent, Canterbury, Kent CT2 7NF United Kingdom}

\cortext[cor1]{Corresponding author}
\begin{abstract}
	Quantum computation represents a threat to many cryptographic protocols in operation today. It has been estimated that by 2035, there will exist a quantum computer capable of breaking the vital cryptographic scheme RSA2048. Blockchain technologies rely on cryptographic protocols for many of their essential sub-routines. Some of these protocols, but not all, are open to quantum attacks. Here we analyze the major blockchain-based cryptocurrencies deployed today---including Bitcoin, Ethereum, Litecoin and ZCash, and determine their risk exposure to quantum attacks. We finish with a comparative analysis of the studied cryptocurrencies and their underlying blockchain technologies and their relative levels of vulnerability to quantum attacks. 
\end{abstract}

\maketitle

\section{Introduction} \label{sec:introduction}

Blockchain systems are unlike other cryptosystems in that they are not \emph{just} meant to protect an information asset. A blockchain is a ledger, and as such it \emph{is} the asset. 

A blockchain is secured through the use of cryptographic techniques. Notably, asymmetric encryption schemes such as RSA or Elliptic Curve (EC) cryptography are used to generate private/public key pairs that protect data assets stored on blockchains. The associated security relies on the difficulty of factoring, when using RSA, or of the discrete logarithm problem with EC.

In a traditional banking system, public- and private-key cryptosystems are used to impose data confidentiality, integrity, and access rules. However, the data itself is decoupled from the key-pair. For instance, if a cryptographic key is lost or compromised, its validity can easily be revoked by a central authority. A new key-pair can be issued and associated to the data. Revoking the key in a timely manner ensures the continued integrity and confidentiality of the data. If a data-breach occurs, servers can be taken offline, and/or backups used. If an account is compromised, often mechanisms exist to allow the legitimate owner to recover this account.

By contrast, in a blockchain system, there is no central authority to manage users' access keys. The owner of a resource is by definition the one holding the private encryption keys. There are no offline backups. The blockchain, an always online cryptographic system, is considered the resource---or at least the authoritative description of it. If a key is lost, this invariably means that the secured data asset is irrevocably lost. If the key, or the device on which it is stored is compromised, or if a vulnerability can be exploited, then the data asset can be irrevocably stolen. In short, in blockchains the protected resources cannot easily be decoupled from the encryption system being used. This makes blockchain technologies particularly vulnerable to advances in quantum technology.

It is infeasible to predict the progress and development of future technology with perfect accuracy. That said, it is possible to extrapolate current and past trends in quantum technology advancement---including all the essential components such as number of qubits, fidelity of gates, error-correction and fault-tolerance\cite{vanmeter}. Doing this, we can confidently conclude that by the year 2035 it is more likely than not that quantum technology will have advanced sufficiently to be able to break RSA2048 efficiently. This conclusion is shared by well established researchers (see, \emph{e.g.}\cite{Aggarwal,mosca}), to the point that the US National Institute of Standards and Technology (NIST) has begun the process of standardizing and deploying quantum-safe public-key cryptography\cite{nist}.

Given the strong coupling between data and cryptosystems in blockchains, the potential vulnerability of these cryptosystems to quantum attacks, the likely introduction of capable quantum computers in the mid-term future---not to mention the usual high monetary value of the assets secured by blockchains---it is important to more deeply understand their current level of vulnerability.

In this paper we analyze some of the most popular blockchain technologies---Bitcoin, Ethereum, Litecoin, Monero and ZCash--- with a particular eye towards their vulnerability to attacks from upcoming quantum technologies. We finish with a comparative analysis of these blockchain technologies, in terms of their relative vulnerability to quantum attacks.

\section{Background}\label{sec:background}

We begin by giving some relevant background information.

\subsection{Quantum Cybersecurity Threats}\label{subsec:quantum}

Quantum computers work by exploiting quantum physical effects to decrease the time required to solve (certain) computational problems by creating and utilizing \emph{quantum superpositions}. 

There are two main families of quantum algorithms that are relevant to the current discussion: subgroup-finding algorithms, and amplitude amplification. 

The first class of algorithms is best represented by Shor's algorithm\cite{shor}. This algorithm can both factor large integers and solve the discrete logarithm in polynomial time. In particular, it can factor an integer $N$ in time $O\left(\log^2 N \log \log N \log \log \log N \right)$ (or more succinctly $O\left(\log^3 N \right)$) and space $O\left(\log N\right)$. Or, as a function of the input size (in bits) $n = \log N$, Shor's algorithm runs in time $O\left(n^2 \log n \log \log n \right)$ (or more succinctly $O\left(n^3 \right)$), using space $O\left( n \right)$.

This is particularly relevant because most public-key cryptosystems deployed today---including RSA, EC, ElGamal and Diffie–Hellman---rely on the computational hardness of either one of these two problems. In order to understand the magnitude of the issue, one can take RSA 2048 as an example. This is considered the `gold standard' for security at the time of writing. A simple calculation shows that it would take a classical computer with a 5Ghz CPU roughly 13.7 billion years to break an RSA 2048 cipher using current best techniques. A quantum computer operating at 10Mhz would be able to do it in roughly 42 minutes\footnote{We calculate this by taking the number of quantum gates---counting error-correction---needed to factor an RSA 2048 public-key.}. In order to do so, however, a device needs to be able to hold in \emph{quantum memory} a state large enough to represent (at least) both the input to the problem, and the output. As discussed earlier, it can be estimated that a quantum computer large enough to break RSA-2048 will likely be ready by the year 2035 (see \emph{e.g.}\cite{Aggarwal,mosca}). 

The second class of algorithm---amplitude amplification\cite{qaa1,qaa2}---consists of generalizations of Grover's search algorithm\cite{grover}. These algorithms allow for a solution to be found in \emph{any} search space of cardinality $N$ in time $O\left(\sqrt{N}\right)$. In short, this allows for any \textbf{NP}-Complete problem to be solved quadratically faster than any known classical algorithm. While the speed-up is a lot less dramatic than in the previous case, the importance of these algorithms rests in their \emph{general applicability}. In short, \emph{any} problem whose solution can be verified efficiently (\textit{i.e.} any problem in \textbf{NP}) admits a quadratic quantum speed-up. Amplitude amplification algorithms are particularly relevant here because many, if not all, consensus algorithms for blockchain technologies rely on solving \textbf{NP}-Complete problems (more details below).

\subsection{Blockchain Technologies}\label{subsec:blockchain}

Blockchain and Distributed Ledger Technology (DLT) markets are predicted to be valued at \$7.59 billion by 2024 \cite{blockchain-value}. Industries that have strong use cases include finance\cite{JP}, logistics\cite{maersk}, and legal fields \cite{raconteur}, with many large global corporations getting on board and integrating the technology: for example IBM \cite{IBM}, JP Morgan\cite{JPMorgan} and Amazon \cite{Amazon}, with Facebook also announcing their own cryptocurrency Libra\cite{Facebook}. This technology removes the need for a trusted third-party to enable the transfer of data and assets. 

Blockchains work on group consensus; the validity of a transaction is determined by a group of nodes that need not trust one another. The blockchain is managed by independent nodes that must reach consensus before updating the ledger with newly validated transactions. There are many mechanisms that enable a network to gain consensus, the most popular being Proof-of-Work (PoW)\cite{bentov2016cryptocurrencies}. This consensus mechanism and underlying cryptographical techniques give blockchains their trustless ability. In general, blockchains work through the linkage of blocks in chronological order. These blocks are groups of transactions of information or cryptocurrency that nodes have broadcast to the network. This forms an immutable series of information, or a chain. Each block in the chain will contain a group of transactions and their information that has been declared to the network. This is generally through the transfer of tokens (cryptocurrency). These tokens hold intrinsic value like traditional fiat currencies---rather than simply hold information about that value like, say, a bank account balance. However, unlike tradition currencies, they are not minted by a central bank. Tokens are distributed to miners, who are nodes that form the group consensus and as such perform work on the network, as a reward for good work. This work primarily consists of creating the blocks in the chain as well as validating that the transactions are well formed and are mathematically fair on the network, \textit{i.e.} not creating or destroying tokens and not spending more than the user transferring tokens can afford. It is through this group consensus that the network and underlying economy of the network can function fairly and independently of any central authority. 

Blockchain technologies can be simplified down to two constituent parts, the consensus protocol and the transaction mechanism. The transaction mechanism is how actors transfer tokens and information; this requires them to provide a digital signature in order to authenticate that they possess the public and private key used to create the digital signature. The consensus mechanism dictates how the verifiers or mining nodes on the network agree on the next blockchain update, which transactions are added, and whether the transactions and the block are cryptographically and structurally valid. 

PoW is the most commonly used consensus mechanism within a blockchain. PoW requires a miner to prove that they have committed a certain amount of effort through the expenditure of computing resources to generate the new block. This mechanism was adopted by Bitcoin forcing the miner, while compiling transactions into a block, to perform some work, \textit{i.e.} spend computational and financial resources to solve a problem. This incentivizes the miner to generate a valid block containing only valid transactions. This work is also easily verified by a any node connected to the network. This expended energy guarantees that a cost is associated with creating a block. Careless or malicious miners that expend the energy to complete a PoW algorithm but have created a \emph{bad block} (a block that includes at least one transaction that if included into the chain would create a wrong state, \emph{e.g.} spending over a user's balance) will be discovered by other nodes in the network. The block will be invalid and this would not be considered by other miners as part of the main chain, leaving the miner financially worse off, as they would receive no mining reward. This ensures the validity of the information contained within the block that is considered by the network as the head of the current longest chain (the block to which miners will attempt to append the next block in the chain). The hardness of this PoW determines how quickly each block is added to the chain: if the hardness of the problem increases then it will take miners longer to solve the problem as it will require more work to be performed by the mining nodes \cite{dificulty}. 

Determination of the ownership of assets within a blockchain network is comparatively more complex when compared to centrally-controlled networks and financial exchanges. A holder of some tokens must be able to demonstrate that they have the ability and authority to spend the tokens. In a centralized system this is kept in check by a central authority. In decentralized systems, cryptographic techniques, such as signature schemes, must be used to demonstrate ownership. 

Signature schemes allow a holder of tokens to cryptographically sign a transaction, and this signature directly relates to the user's public/private key pair from which their account is created. There are many different signature schemes, for example ElGamal\cite{elgamal}, RSA\cite{RSA}, and Schnorr \cite{schnorr}. Elliptic Curve Digital Signature Algorithm (ECDSA) is a signature scheme that relies on the difficulty of solving the discrete logarithm problem over elliptic curves. Compared to other schemes, ECDSA allows for the same level security using smaller keys---for instance, 256 bits compared to 2048 bits in the case of RSA \cite{levy}. Signing and verification speed as well as compact signature size are all essential for blockchain technologies. This makes ECDSA particularly well-suited for use in blockchains. However, given its reliance on the discrete logarithm problem for its security, coupled with its smaller key-sizes, ECDSA is particularly vulnerable to quantum attacks\cite{donny}.

\section{Related Work}\label{sec:related}

The current literature covers security analysis of all major cryptocurrencies\cite{lim2014analysis, praitheeshan2019security, kappos2018empirical, moser2017empirical}. This includes all the blockchain cryptocurrencies covered here, as well as third party elements to a blockchain's infrastructure \emph{e.g.} cryptocurrency wallets and client providers \cite{sai2019privacy, he2020security}. All this said, the literature so far has focused almost entirely on cybersecurity threats from \emph{classical} actors, and has almost entirely ignored the growing threat from quantum attacks.

In the wider more general field of cryptography there is a comprehensive study of quantum attacks and methods for protecting against them. This area of study is called \emph{post-quantum cryptography}, and it is too vast to properly cover here, but a few good resources include\cite{lohachab2020comprehensive, nejatollahi2019post,perlner2009quantum}. However, the field of \emph{post-quantum blockchain cryptography} seems to be fairly barren. The existence of a quantum advantage applicable to blockchain technologies has certainly been noted before see \emph{e.g.}\cite{cui2020threats, suo2020quantum, li2020efficient, wu548quantum}. 

This realization has led to work on blockchain systems that are more resilient against quantum attacks (see \emph{e.g.} \cite{li2020efficient, wu548quantum}), as well as blockchain technologies such as Corda \cite{corda_2020}, Bitcoin Post-Quantum \cite{anhao2018bitcoin} and Abelian \cite{abelian} that are seeking to provide post-quantum infrastructure to the blockchain sector. However, these projects are either in an early development stage or simply not widely used as of this writing. There are no known plans to incorporate this type of work into existing, widely-used, cryptocurrencies.

To our knowledge, the only post-quantum, in-depth, rigorous analysis of a blockchain, before this paper, is the work by Aggarwal \emph{et. al.} on the security of Bitcoin to quantum attacks\cite{Aggarwal}. To the best of our knowledge, the work presented here is the first attempt at a more comprehensive, rigorous study of the vulnerabilities of cryptocurrencies to quantum attacks.

In this paper we consider five major cryptocurrencies (and some of their variants): Bitcoin, Ethereum, Litecoin, Monero, and ZCash. Excepting Bitcoin, none of these cryptocurrencies have a (publicly available) rigorous post-quantum vulnerability analysis. We can, however, consider the current state of the scientific literature on \emph{classical} cybersecurity attacks, to provide some further context for the work presented here.

Bitcoin being the oldest and most popular cryptocurrency has been analyzed extensively from a classical computing perspective. The analysis of Bitcoin covers both the underlying protocol \cite{garay2015bitcoin} as well as the cryptography which secures transactions \cite{giechaskiel2016bitcoin}. Furthermore, Bitcoin is the only blockchain that has had analysis against quantum attack performed on it \cite{Aggarwal}. 

Litecoin as a hard fork of the Bitcoin blockchain shares a majority of its infrastructure, and while there has not been as much dedicated research towards the project as Bitcoin, analysis of its protocol structure is well documented from a security perspective \cite{popuri2016empirical, wang2018overview}. Research into the security of the Equihash PoW algorithm used by Litecoin has also been performed \cite{alcock2017note}. 

Ethereum, unlike Litecoin, does not share many similarities with Bitcoin as it is its own unique protocol (and does not hardfork from any other blockchain). Much of the security analysis of Ethereum is based on its novel blockchain feature: \emph{smart contracts}\cite{mense2018security, wohrer2018smart}. Due to Ethereum's differences to other protocols, there is also extensive research analyzing its security from classical attacks \cite{chen2020survey, ma2021security}. 

Both Monero and ZCash, in comparison to the other blockchains analyzed here, have a smaller user base. Despite this fact, significant security analysis has been performed on both these cryptocurrencies due to their unique usage of confidential transactions \cite{sun2017ringct, wijaya2018monero, chervinski2019floodxmr, kappos2018empirical}. 

A significant amount of literature does not neatly fit into the categories above since it covers the security of various blockchains at once, see \emph{e.g.} \cite{homoliak2019security, li2020survey, zhang2019security}. This type of analysis makes sense because of the similarities between blockchains as well as the similarities of the structure of their cryptographic protocols. Generally, most blockchains use ECDSA (or some variant of the scheme) in order to provide cryptographic signatures to prove ownership over assets on the blockchain, and PoW remains the most popular mechanism for generating consensus on blockchain networks. While from a classical perspective many of the small differences in the protocols have little impact on the overall security of the network, these differences can have a significant impact on how severe a quantum attack will be on the network, as we show in the subsequent sections.

\section{Methodology}\label{sec:Methodology}

First, we selected a representative set of blockchain technologies. We considered several factors in deciding whether to include a particular blockchain technology. The first is popularity, measured by popular interest using Google Trends, and academic interest using number of academic citations. We also strived to ensure a diverse set in terms of technological and cryptographic techniques. For each blockchain technology selected, we carefully studied the cryptographic primitives used, and their level of reliance on cryptographic protocols known to be vulnerable to quantum attacks. This analysis is in-line with the original analysis of Bitcoin done by Aggarwal \textit{et. al.}\cite{Aggarwal}. For each technology, we considered two primary attack vectors. First, we consider attacks against the blockchain network's consensus mechanism using a quantum amplitude amplification algorithm. Then, we study Shor's algorithm-based attacks against the blockchain scheme's transaction signature schemes. When relevant to a blockchain technology, we also study potential attack vectors not covered by Aggarwal \textit{et. al.}. Additionally, considerations such as the attractiveness, or profitability of an attack are discussed when applicable.

Having completed the analysis described above, we collate the results in the following way. For each selected blockchain technology, we present and rank its vulnerabilities to quantum attacks. We then describe the most damaging attack in terms of potential financial or reputation loss to the network. Whether these vulnerabilities can be removed or mitigated is also considered where relevant. These factors are then combined into a score in Table\ \ref{table} that represents the blockchain's overall vulnerability. This allows us to rank blockchains by their relative vulnerability from those with a limited potential for quantum attack given a vulnerability score of low, to a blockchain that could be rendered completely unfit for use by the introduction of quantum technologies given a vulnerability rating of \emph{very high}. On the other end, a blockchain that is susceptible to quantum attack, but employs technologies that could dissuade an attacker or make an attack more difficult, would be given a rating of \emph{medium}. This information is summarized in Table\ \ref{table}, on Page\ \pageref{table}.

Finally, the estimate that we give of the year 2035 for the likely introduction of quantum computers that can break RSA 2028 is based an extrapolation of current and past trends in the essential components of quantum technologies such as number of qubits, fidelity of gates, error-correction and fault-tolerance\cite{vanmeter}. We base our estimate on a consensus of various experts in the field\cite{mosca,Aggarwal}, as reflected by official state policy\cite{nist}.

\section{Results}\label{sec:Results}

In this section we discuss several blockchain technologies, the cryptographic schemes they use, and how these dependencies can be exploited by a quantum-capable attacker. While the work in this section is almost entirely original, the results in the Bitcoin section follow Aggarwal \textit{et. al.}\cite{Aggarwal}, who first reported these findings.
\begin{table}
	[h!] \centering \label{table2} 
	\begin{tabular}
		{|l|p{50pt}|p{50pt}|} \hline \textbf{Blockchain}& \textbf{Subgroup-Finding algorithm (Shor's)} & \textbf{Amplitude Amplification (Grover's)} \\
		\hline Bitcoin & \cellcolor{red}\xmark & \cellcolor{orange}-- \\
		\hline Ethereum & \cellcolor{red}\xmark & \cellcolor{orange}-- \\
		\hline Litecoin & \cellcolor{red}\xmark & \cellcolor{orange}-- \\
		\hline Monero & \cellcolor{red}\xmark & \cellcolor{green}\checkmark \\
		\hline ZCash & \cellcolor{red}\xmark & \cellcolor{orange}-- \\
		\hline 
	\end{tabular}
	\caption[Vulnerabilities of Key Blockchain Technologies]{ \textbf{Vulnerabilities of Key Blockchain Technologies:} This table shows the vulnerabilities of key cryptocurrencies against two forms of quantum attack. An \xmark denotes the blockchain has strong vulnerabilities against quantum attacks: due to the exponential quantum advantage for such attacks, as soon as quantum computers exist with sufficient memory, these could be used to effectively attack the blockchain in question. A -- denotes that the blockchain has an intermediate level of vulnerability: while a quantum advantage exists, this is only quadratic in nature, hence it will take longer for quantum technologies to advance to the point of becoming a threat. Finally, a \checkmark means that the cryptocurrency is currently considered safe from quantum attacks. } 
\end{table}

\subsection{Bitcoin}

Bitcoin, first described in a paper by a person or a group under the pseudonym Satoshi Nakamoto \cite{Bitcoin}, is the most popular and first true blockchain technology. The 2008 paper paved the way for the development of the distributed ledger technological space. Designed as a peer to peer payment method, it removed the need for a central authority. It is underpinned by cryptographic schemes that allow peers in the network to validate transactions in a trustless environment, and store these in a ledger that is cryptographically secure and immutable. These cryptographic techniques are secure from attack on a classical computer, however, they can be exploited by a sufficiently powerful quantum computer. 

Bitcoin uses Hashcash as its PoW mechanism. Hashcash\cite{hashcash} was originally designed as a denial of service countermeasure for email systems. This was done by requiring the potential sender to expend time solving a computationally hard problem, before being able to send an email. As implemented in Bitcoin, Hashcash requires the prospective miner to calculate a SHA-256 hash value for the header---plus some random number--- so that the hash value is smaller than a predetermined number. This number is an adjustable parameter in the Bitcoin network. The smaller the number, the higher the computational difficulty of the problem.

The use of the Hashcash PoW mechanism has two net effects. One, with the current high difficulty parameter, miners are incentivized to use specialist hardware, such as ASIC miners, and/or join mining pools where the work and reward are divided among various users. More importantly, PoW de-incentivizes attempts to add bad blocks to the networks. These blocks have a very high chance of being rejected by the network due to error correction, and therefore ensuring very high wasted costs on the potential miner.

The transaction mechanism within Bitcoin uses ECDSA signature scheme in order to prove authority and ownership of tokens, as well as irrefutable evidence that the tokens have been spent and that the transaction has not been meddled with after the transaction has been signed. The elliptic-curve Bitcoin employs for its ECDSA is secp256-k1. The signature in Bitcoin is made up of two values $S$ and $R$ \cite{antonopolous}. $R$ is the $x$ coordinate of a point of an elliptic curve\footnote{A further description of elliptic-curve cryptography can be found here\cite{ecc}.}. This point is the public key of an ephemeral public / private key pair, created by a user during the process of signing the transaction. $S$, the other half of the signature can then be created as follows: $S = k^{-1} (SHA256(m) + dA \cdot R)~mod~p$. Where $k$ is the temporary private key, $SHA256(m)$ is the hash of the transaction message, $dA$ is the signing private key, and $p$ is the prime order of the elliptic field. Given $S$ and $R$ of a signature, the signature can be validated by any user as $K$, the ephemeral public key, can be found from the two parts of the signature. During this process a user must also declare the public key associated with their account in order for validation to occur. This process signature is required for every input of a transaction, otherwise the transaction will be invalid and will not be added in a block in the blockchain. 

Bitcoin and its underlying cryptographic schemes are vulnerable to possible quantum attacks. One such attack uses Grover's search algorithm to perform PoW at a much faster rate than classical miners. An attacker would aim to generate just over as much PoW as the rest of the network combined---effectively forcing consensus on any block the attacker so desires (this is known as a 51\% attack). 

Current ASIC miners are capable of performing roughly 18TH/s. This, combined with the current sizes of Bitcoin networks, makes a quantum based 51\% attack infeasible for the time being. Our own calculations based on current ASIC technology, as well as that of other authors\cite{mosca,Aggarwal}, put the earliest likely date that this type of attack will be possible at 2028. However, advances in ASIC technology are likely to push back this date much farther.

However, the most damaging attack on the Bitcoin blockchain is on its ECDSA scheme; more specifically upon transactions that had been declared to the network and not yet added to the blockchain. The hardness of ECDSA relies on the hardness of the Elliptic Curve Discrete Logarithm Problem. As noted in the previous section, this problem can be solved in polynomial time $O(n^3)$ on a quantum computer of sufficient size where on a classical computer it can be solved in exponential time at approximately $O(2^n)$. While on the bitcoin network it could be possible to not only hijack individual transaction it must also be considered that a quantum attack could also aim to take control of a users entire bitcoin wallet. If the same public / private key pair is used to hold the users bitcoin after the public key becomes public knowledge, then all funds secured by the key pair will be vulnerable. However, it must also be considered that bitcoin wallets tend to not repeatedly use the same key pairs. Bitcoin transactions send an entire UTXO (potentially multiple depending on whether the user is in possession of one UTXO that is greater than the amount to be transacted). In the most simple form of bitcoin transaction, one UTXO will be spent as the input, there will then be two outputs. One creates a new UTXO of the amount being transferred to the relevant account. If the UTXO in the input is greater than the UTXO in the output a second UTXO is created as the change, which will be returned to the original user. This account where the change is sent to typically will be controlled by a newly generated public / private key pair. This means that an attack designed to get access to the entirety of the users wallet while possible on the bitcoin network is less likely by the common security mechanism of changing the public / private key pair after every transaction. 

An effective quantum attack would consist of finding the private key when the public key is revealed following the broadcast of a signed transaction to the network. This would allow an attacker to sign a new transaction using the private key, thus impersonating the key owner. As long as the quantum attacker can ensure that their transaction is placed on the blockchain before the genuine transaction, they can essentially `steal' the transaction and direct the newly created Unspent Transaction Output (UTXO) into whichever account they choose. It is easy to calculate that a quantum computer with 485550 qubits and running at a clock-speed of 10GHz could solve the problem using Shor’s algorithm in 30 minutes\cite{Aggarwal}. At the same time, the average waiting time for transactions in the pool of transactions currently waiting to be integrated into a block in the Bitcoin blockchain frequently exceeds 30 minutes \cite{blocktime}. This makes this type of attack quite feasible.

Early in the implementation of Bitcoin it was possible for Bitcoin users to be paid directly to their public key (P2PK), rather than to the hash of the public key, which is commonly known as the user's payment address. This has potential repercussions for older Bitcoin accounts. An example of this is the first ever Bitcoin transaction between Hal Finey and Satoshi Nakamoto at block 170 of the Bitcoin blockchain. This form of transaction is common in early coinbase transactions used to reward Bitcoin miners. This means that some of the original (often quite affluent) accounts may have revealed their public key in the early stages of the Bitcoin blockchain. 

Thus, these accounts are extremely vulnerable to quantum attacks using Shor's algorithm. Unlike the attack on the ECDSA signature scheme described earlier, there is no time limit to perform this type of attack. Once a sufficiently large quantum computer exists (estimated by the year 2035), a quantum attacker can easily calculate these accounts' private keys, sign new transactions as these users, and empty these accounts of all their funds.

The threat of quantum attack has given rise to a project called Bitcoin Post-Quantum. \cite{anhao2018bitcoin} This project hard forked from the bitcoin network at block height 555,000 (mined on 22/12/2018). This project utilizes a quantum secure digital signature scheme \cite{hulsing2018xmss} as well as implementing a Proof-of-Work mechanism utilizing the birthday paradox as in Z-Cash discussed in section \ref{sec:zcash}. Because this project is a fork, however, it provides no actual security benefits to the original Bitcoin chain. Hence, the discussion in this section is still very much relevant.

In summary, Bitcoin will be very vulnerable to quantum attacks using Shor's algorithm. The most wide-spread vulnerability open to attack will be transactions that have been declared to the network and not yet added to a block. The most vulnerable accounts are those that divulged their public-key in the earlier days of the Bitcoin network. Finally, Bitcoin's consensus mechanism exhibits a vulnerability to Grover algorithm-based attacks. However, since Grover's algorithm only provides a quadratic advantage, advances in classical computer technology are likely to keep Bitcoin secure against this type of attack for much longer than for Shor's algorithm-based attacks. 

\subsection{Ethereum}

Ethereum is considered the second generation of blockchain technologies. It has an associated cryptocurrency, Ether, and introduced the use of smart contracts and distributed Applications(dApps). Ethereum uses an account-based system where each transaction will deduct or add Ether to a user's account. Smart contracts allow users of the blockchain to create a computationally-binding contract, meaning that it allows the creation of transactions dependent on certain trackable objectives. 

Ethereum is currently transitioning from a Proof-of-Work (PoW) consensus mechanism, to a Proof-of-Stake (PoS) one. EthHash is a PoW mechanism that is used in the current implementation of Ethereum at the time of writing this paper. A single round of SHA-3 (Keccak-265) hashing is used to create the PoW problem, in a similar manner to Bitcoin. Mining nodes then compete to generate a hash that solves the PoW problem. The second is the currently not implemented PoS mechanism known as Casper \cite{casper}. As mentioned earlier, PoW is used to provide a computationally tasking problem to ensure that a block produced by a miner is valid. PoS however dissuades bad miners from attempting to subvert the system as they risk losing their Ether if they perform a poor job. The security in the system relies on the fact that the larger the stake the more voting power a miner will receive, by staking more coins a user is more likely to behave honestly as they have more to lose if discovered. Further security is gained from the disincentive that a user would have to own a large amount of Ether in order to perform an attack, a successful attack would inevitably cause price drops for the cryptocurrency thereby negatively impacting a user that is wealthy in ether. 

Ethereum, like Bitcoin, uses a variant of the ECDSA scheme based on the secp256-k1 elliptic curve\cite{ant-eth,wood2014ethereum}. In an Ethereum transaction there is no `from' field, which means that the primary public key $K$ associated with account is not explicitly revealed. It can however be retrieved through a process called public key recovery where a user can reconstruct the public key from another user's transaction signature. 

Similar to Bitcoin's consensus mechanism, a quantum attacker can make use of Grover's algorithm to attack EthHash. We can calculate the hash rate possible on a quantum computer against the Ethereum as follows. First, we calculate the difficulty $D$ of the PoW for Ethereum: $D = \frac{H_r \times B}{2^{32}}$ where $H_r$ is the network hash rate and $B$ is the block time of the blockchain. In Ethereum $B$ is currently 16 seconds, while $H_r$ is currently $18 \times 10^{13}$ H/s \cite{EthHash}. Therefore, the difficulty value is currently 670552. The equivalent hash rate for a quantum computer is $h_q = 0.04 \times s \sqrt{D}$, where $s$ is the clock speed of the computer. Even without any advances in ASIC technology, a quantum attacker would require a clock speed of about 5THz before being able to attempt a 51\% attack Ethereum's consensus mechanism.

The Ethereum signature scheme is highly insecure against attacks using Shor's algorithm, since Ethereum's signature scheme relies on the hardness of the discrete logarithm problem. This can be solver in polynomial time ($O(n^3)$) using Shor's algorithm compared to exponential time ($O(2^n)$) on classical infrastructures. Ethereum does have one minor advantage in that it has a significantly shorter transaction processing time when compared to Bitcoin. This is countered, however, with one major disadvantage: Ethereum's use of account-based transactions. Every single outgoing transaction needs to be signed using the account's private key, and can be verified using the public key. Once a user has an outgoing transaction, the account's public key is available to anyone that reconstructs it using the key recovery process mentioned earlier. A quantum assailant can thus request the public key, calculate the private key using Shor's algorithm, and thus takeover the entire account. This vulnerability is exacerbated by the existence of tools such as Etherscan \cite{etherscan} that allow an assailant to search for, and target, accounts holding a large amount of Ether. This attack would be severely damaging as the potential reward (for the attacker) and loss (for the victim) would be significantly higher when compared to targeting individual transactions since the quantum attacker would be targeting an entire accounts balance of tokens. 

In summary, while Ethereum has a considerably shorter block-time when compared to Bitcoin it is significantly more vulnerable to quantum attack due to its account-based transaction system. While some other blockchains allow a user to reuse the same public key for multiple transactions, it is far less common and users are dissuaded from this practice. In Ethereum, all outgoing transactions are signed using a single private/public key pair associated with the account. This makes the entire account balance vulnerable after a single outgoing transaction.

\subsection{Litecoin}

Litecoin is a source-code fork of the Bitcoin blockchain. This means that it shares many similarities with Bitcoin. However, Litecoin also has marked differences: these include the block time as well as the PoW mechanism\cite{litecoin}. It has very similar use-case to Bitcoin as an electronic payment method. However, due to a shorter block time, its goal is to process transactions faster than Bitcoin. 

Litecoin uses a different PoW scheme than Bitcoin, called Scrypt. It has the same goal of expending computing resources in order to solve a problem to give a user authority to create the next block on the chain. Scrypt is designed to use significantly less hashing power; this can be seen in comparison with Bitcoin where the hashing rate is approximately 46,000,000TH/s \cite{blockchain-dot-com} against 298TH/s \cite{bitinfo} for Litecoin. 

Scrypt is a simplified version of the password derivation function created by C. Percival \cite{percival}, originally for the Tarsnap online backup system. Scrypt differs from other PoW schemes in that rather than being highly intensive on the processing power, it is highly intensive on the use of RAM on the mining node. This originally was chosen in order to reduce the advantage of using---and hence prevalence of---ASIC miners when compared with blockchain technologies. However it was proven relatively quickly that Scrypt was not ASIC-resistant \cite{litecoinminer}. 

Litecoin uses an ECDSA scheme in order to sign transactions. Similarly to Bitcoin, it implements its signature scheme using the secp-256k1 elliptic curve. 

Like other PoW systems, Scrypt is potentially vulnerable to a quantum 51\% attack using Grover's algorithm. Litecoin's current hash rate is 320TH/s\cite{litehash}. Litecoin's difficulty can be calculated as: $D = \frac{32 \times 10^{13} \times 150}{2^{32}} = 11175870$. Thus a quantum computer would have to run at a clock speed of 2.4 Thz to even attempt such an attack at current hash rates. This, plus future improvements in ASIC technology make this type of attacks unlikely in the foreseeable future.

Because of its use of ECDSA, Litecoin is vulnerable to quantum attacks in polynomial time of $O(n^3)$ while using Shor's algorithm performed against transactions that are awaiting to be incorporated into a block. This is likely to be the most profitable attack for a quantum attacker. Litecoin has the advantage of a shorter block time and a slightly quicker throughput when compared to Bitcoin. Therefore, Litecoin has some minor improved resistance against quantum attacks when compared to Bitcoin. This advantage is however minimal: given a quantum computer capable of attacking Bitcoin, a slight increase in its clock speed would suffice to make it capable of attacking Litecoin.

While this section focused on Litecoin, a similar analysis applies to many more `altcoins' that are based on the Bitcoin blockchain or the original Bitcoin code. These range from direct hard forks of the Bitcoin blockchain of which there are 45 current active projects, through Bitcoin cashcite{bcash}, Bitcoin gold\cite{bgold} and Bitcoin core\cite{bcore}, to source code forks like Litecoin. While a detailed discussion of each of every single \emph{`altcoin'} is necessarily beyond the scope of this paper, this section serves to highlight the vulnerability of all ECDSA based blockchains---which includes almost all Bitcoin forks---to quantum attacks that use Shor's algorithm.

In summary, due to its similarities to Bitcoin, Litecoin displays the same vulnerabilities to quantum attacks. Moreover, Litecoin can be used to demonstrate the severe vulnerabilities faced by blockchain technologies based on Bitcoin. 

Many of these altcoins have significantly lower transaction processing times than Bitcoin. This gives these blockchains slightly higher resilience to Shor algorithm-based attacks---though they are all ultimately quite vulnerable to such attacks.

On the other hand, given current hash rates, and likely improvements in ASIC technology, Litecoin is likely to be safe from Grover's algorithm-based attacks on its consensus mechanisms for the foreseeable future. However, a drop in this hash rate---for example, due to a reduction of the block reward for completing the PoW as has happened before\cite{litecoinhashrate}---could leave the network more vulnerable. 

\subsection{Monero} \label{sec:monero} Monero is a blockchain that focuses on the privacy of its users. A majority of blockchains advocate anonymity through the use of pseudonyms. Pseudonym identities however do not provide a user with anonymity as their pseudonym is known to other users. Through the use of chain analysis techniques it is possible to discover who has sent and received transactions, furthermore the number of tokens sent or received, or account balances. Monero provides obfuscation of both a user's identity and value of transactions through the use of further cryptographical techniques. It offers true anonymity to its users through the use of Pedersen Commitments \cite{Pederson} and Range Proofs \cite{Range}. 

Monero uses the ASIC-resistant CryptoNight v8 PoW scheme which is derived from the Egalitarian Proof of Work from CryptoNote \cite{cryptonote}. The scheme relies on access to slow memory at random intervals. CryptoNight is particularly memory intensive, requiring 2Mb per instance. 

EdDSA is used as the signing algorithm in Monero. EdDSA is implemented using the twisted Edwards curve Ed25519. This signature scheme is a variant of ECDSA and is still reliant on the hardness of the discrete logarithm problem. A keccak-256 (SHA-3) hashing function $\mathbb{H}$ is used. The signature for signing a transaction using EdDSA is made up of two parts $R$ and $s$ \cite{monero}. First, a user must compute the hash of their private key $k$ so that $\mathbb{H}(k)$ to create $h_k$. They then compute $r = \mathbb{H}(h_k, m)$ where $m$ is the message of the transaction. $r$ is then associated with a generator of the elliptic curve $G$ to form $R = rG$. The second signature component $s$ is then computed as $s = (r + \mathbb{H}(R,K,m)) \cdot k$, where $K$ is the user's public key. This signature scheme is extended in Monero, through the use of ring signatures. 

A further area of interest in Monero is how it gains transaction anonymity. It does so through the use of three technologies working together: stealth addresses, ring signatures and ring confidential transactions. 

In simple terms, stealth addresses and ring signatures work in the following way. For every transaction, Monero also broadcasts several `fake' inputs to the transaction. Only the senders and receivers of the transaction will know which is the correct commitment for the transaction, as the senders and receivers of tokens share a secret key. Moreover, if there are multiple recipients within the transaction, only the sender will have knowledge of the whole transaction. The process consists of the user including one input using UTXO (balance) from their wallet, and padding with extra randomly selected spent outputs to the transaction up to the ring size. For instance, if the ring size is five then a further four randomly selected spent outputs are added as inputs into the transaction. Which input is the correct one (signed by the user) will not be deducible to other users\cite{RingCT}.

The Monero network needs a way to ensure that the above transactions balance correctly, in other words that the incoming currency into the transaction equals the outgoing currency. Monero's current mechanism for doing so is called Bulletproof,\cite{bulletproof}. Bulletproof is a zero knowledge proof protocol that can ensure the balance of transactions. It is much more efficient than previous zero knowledge range proofs, both in computational terms and the amount of space required on the blockchain to record these proofs.

Monero very recently moved it's PoW scheme from CryptoNight to RandomX \cite{randomx}. RandomX is PoW system based on the execution of random programs in a special instruction-set that consists of integer math, floating point math and branches. This PoW system was developed with the intent of minimizing GPU advantage in PoW. However, it is possible that this may also, indirectly, lead to more quantum resiliency. As of this writing, no method for gaining quantum advantage for RandomX is known.

Monero's signing algorithm EdDSA, like ECDSA, relies on the hardness of the discrete logarithm problem for its security, making it highly susceptible to quantum attacks using Shor's algorithm in $O(n^3)$ computations. However, Monero's privacy system gives it some added level of security. An attacker would not know the amount being transferred in a target transaction. Hence, transactions of value are unobservable without prior attacks. Further, the use of RingCT means that the quantum assailant would need to solve multiple Pedersen commitments in order to find the correct public key used in the transaction. This makes Monero slightly more secure against---or at least a slightly less attractive target for---quantum assailants than other blockchain networks.

Bulletproofs are particularly susceptible to quantum attack. They rely on the discrete logarithm problem for their hardness and so similarly can be solved in polynomial time of $O(n^3)$. The security relies on the fact that no-one knows any $xG = H$ and no $xH = G$ for the Pedersen commitment. A quantum attacker could breach the commitment revealing the values contained within. This would allow the attacker to reveal all previous transactions that have been obfuscated, since one of the key features of a blockchain is that it is immutable. While this does not have any financial benefit, the information gained could be valuable, as the hidden information may be confidential, and could potentially be used to extort users of the network.

In summary, Monero transactions are highly vulnerable to quantum attacks---though the network's transaction anonymization makes these less attractive targets for attack than transactions in other blockchain networks. However, it should also be noted, Monero's PoW system---RandomX---is the only such system with no known quantum vulnerabilities. 

\subsubsection*{Beam and Grin}

Beam\cite{beam} and Grin\cite{grin} are similar to Monero in that they use Pedersen commitments to mask the amounts transferred. However, they use a technique called Mimblewimble. Mimblewimble is an obfuscation protocol like Bulletproof. Here, each newly created UTXO is obfuscated by a blinding factor. This blinding factor hides the amount represented by the UTXO and this provides an extra level of anonymity to the blockchain. \cite{mimblewimble}. 

Like Monero, both Beam and Grin are vulnerable to quantum attacks against both their obfuscation technique as well as their signature scheme. Thereby, attacks presented against Monero are equally valid against these two blockchains. However, as with Monero, the obfuscation of account and transaction values provides both of these blockchains with an element of resilience. While the obfuscation can be removed by a quantum attacker, the attacker has no way of knowing whether the transactions that they are attempting to view are of significant enough value to warrant performing a quantum attack. 

\subsection{Zcash} \label{sec:zcash} Zcash is another privacy-based blockchain. Unlike Monero, however, Zcash allows for transactions to go from private accounts to public ones, and \emph{vice versa}. Anonymity is integral to the Zcash blockchain. Rather than pseudonymising the identity of users through the use of account address, input and outputs can be obfuscated. It allows private transactions as well as public transactions. Zcash transactions implement zero-knowledge proofs in the form of zk-snarks (Zero-Knowledge Succinct Non-Interactive Argument of Knowledge); these use a trusted set up. Trusted setup uses some form of publicly-available element as part of the proving mechanism for a transaction. These publicly-available elements are either generated by a central entity, or alternatively in collaboration with the entire network in the form of a public ceremony \cite{public}. Zk-snarks are a zero-knowledge proof system that allows a user to demonstrate that a transaction they are sending is fair while not revealing the amount being transacted. Within Zcash there are four transaction types \cite{zcash}: private, deshielding, shielding and public. Private transactions obfuscate the amount being transacted at input and output. Shielding transactions obfuscate previously publicly-visible transactions, while de-shielding does the opposite. Public transactions can be considered ``traditional transactions", similar to those in other blockchains, in that they employ only pseudonym identities to protect users, and in that the value being transacted is publicly visible. 

Zcash uses the Equihash PoW to gain consensus. Equihash \cite{equihash} is a memory-hard PoW based on the generalized birthday problem. Equihash has the parameters $n$ and $k$. $k$ is the target value while the miner is given a sequence of $X_{1...N}$ $n$ bit strings. The miner must find $2^k$ distinct $X_{ij}$ such that $\oplus^{2^{k}}_{j=1} X_{ij} = 0$. The solution to this problem is found using Wagner's algorithm, which is the most efficient known algorithm for finding the solution to this problem on a classical computer. 

ZCash implements EdDSA, instantiated on the Ed25519 curve. A signature consists of two parts S and R. This is defined by the scheme described in \cite{bern} and is also described in the ZCash white paper \cite{zcash}. This signature scheme is reliant on the hardness of the discrete logarithm problem. The signature consists of: 
\begin{itemize}
	\item An elliptic curve generator $B$ of mod $\mathcal{L}$ 
	\item A cryptographic hash function $H$ which in the case of ZCash is BLAKE-2b-256. 
	\item $M$ which the message being signed 
	\item The private key $a$ 
	\item The public key $A$ which is generated from the private key in the form $A = aB$ where $B$ is a generator on the Ed25519 elliptic curve. 
\end{itemize}
First $r$ is created where $r = H(a,M)$, then $r$ is multiplied by $B$ so that it is $R = rB$. $S$ can then be computed as $S = (r + H(R, A ,M)a) \text{ mod }\mathcal{L}$, giving the signature $Sig = (S, R)$.

Zk-Snarks, as previously stated, rely on a trusted set-up. This set-up requires the pre-generation of a \emph{greater public-key}. This global public-key must have no private-key, otherwise the holder of such a key could create ZCash tokens at will. This greater public-key is created by many users collaboratively creating shards, or small portions, of a greater public-key from individual user private-keys. After the public ceremony, if at least one user destroys their individual private key, then attempting to find the greater public-key's corresponding private-key becomes computationally infeasible on a classical computer. This is because this private-key can only be computed by either solving the discrete logarithm problem, or by having access to all the ephemeral private keys used to create the greater public-key.
\begin{table*}
	[!ht] { \label{table} \tiny 
	\setlength{\tabcolsep}{3pt}
	
	\begin{tabularx}
		{
		\textwidth}{|l|l|p{45pt}|X|}
		
		\hline \textbf{Blockchain}& \textbf{Risk Level} & \textbf{Target} & \textbf{Vulnerabilities}\\
		\hline \textbf{Bitcoin}& High& Transactions declared to the network & Transactions declared to the network are vulnerable to quantum attack, specifically with regards to their signature scheme. The main form of attack identified is against transactions declared to the network which have not yet been incorporated into a block. Using the public key declared by the sender of a transaction, a quantum attacker can find the private key. This will allow them to duplicate the transaction with whichever output location they desire. \\
		\hline \textbf{Ethereum}& High& Re-use of public keys & Ethereum is designed on an account-based system, within which reuse of public keys is common. The attack mechanism we have identified can target accounts that have previously declared transactions to the network, while still retaining some Ether tokens in the account. By solving the public key to gain the private key using Shor's algorithm, a quantum attacker could forge transactions in a user's name, by generating a valid transaction signature. \\
		\hline \textbf{Litecoin} & High & Transactions declared to the network & As Litecoin shares a majority of its technical structure with Bitcoin, it is equally vulnerable to quantum attack. The most damaging attack technique as in Bitcoin is against transactions declared to the network that have not yet been added to the blockchain. \\
		\hline \textbf{Bitcoin Gold} & High & Transactions declared to the network & Due to the similarities with the Bitcoin cryptographic elements, Bitcoin Gold shares the same vulnerabilities.\\
		\hline \textbf{Bitcoin Core} & High & Transactions declared to the network & Due to the similarities with the Bitcoin cryptographic elements, Bitcoin Core shares the same vulnerabilities.\\
		\hline \textbf{Bitcoin Cash} & High & Transactions declared to the network & Due to the similarities with the Bitcoin cryptographic elements, Bitcoin Cash shares the same vulnerabilities.\\
		\hline \textbf{Monero}& Medium& Obfuscated transactions and transactions declared to the network & The signature scheme used in Monero EdDSA is vulnerable to quantum attack as it relies on the discrete logarithm problem. However Monero gains some resilience to quantum attack through the anonymity of its users as well as the amounts being transacted. Although the Bulletproof protocol used in Monero to achieve this obfuscation of transacted amounts is vulnerable to quantum attack, an attacker would be reliant on luck in order to select a transaction of significant value. Furthermore, due to a recent change in the consensus protocol implemented on Monero where RandomX was introduced, it would also have further resistance to quantum attacks attempting to perform a 51\% attack utilizing Grover's algorithm. \\
		\hline \textbf{BEAM}& Medium& Obfuscated transactions and transactions declared to the network & BEAM's signature scheme, as well as the obfuscation technique Mimblewimble, are vulnerable to quantum attack. Quantum attack could both, intercept transactions broadcast to the network and remove anonymity from hidden transactions. However as with Monero, the hiding of transaction and account values removes some of the incentive for a quantum attacker. \\
		\hline \textbf{Grin}& Medium& Obfuscated transactions and transactions declared to the network & Grin's signature scheme, as well as the obfuscation technique Mimblewimble, are vulnerable to quantum attack. Quantum attack could both, intercept transactions broadcast to the network and remove anonymity from hidden transactions. However as with Monero, the hiding of transaction and account values removes some of the incentive for a quantum attacker. \\
		\hline \textbf{ZCash} & Very High & Public parameter generated during the Zk-SNARK ceremony & ZCash is highly vulnerable to quantum attack against both its consensus algorithm and its signature scheme. However, the most damaging attack found against ZCash is vulnerability of its zero-knowledge proof protocol ZK-SNARKS, as this obfuscation method requires a trusted set up and therefore the production of a public parameter, which is a public key. If a quantum attacker gains the private key to this public parameter, they will be able to generate tokens at will. \\
		\hline 
	\end{tabularx}
	} 
	\caption[Blockchain Quantum Vulnerability Overview ]{ \textbf{Blockchain Quantum Vulnerability Overview:} This table shows a summary of the blockchain vulnerabilities discussed in this paper. The table shows, from left to right, the blockchain in question, the level of risk established here, the particular underlying cryptographic technology at risk, and a summary of the attack. } 
\end{table*}

Zcash is open to quantum attacks in three distinct ways. The first one is quantum attacks against its consensus mechanism. Grassi \cite{Grasi} \textit{et. al.} developed a quantum algorithm for the $k-xor$ problem (generalized birthday problem), that improves on Wagner's classical algorithm. This quantum algorithm has an improved time and memory complexity of $O\left(2^{n/(2+\lfloor log_2(k)\rfloor)}\right)$ compared to Wagner's $O\left(2^{n/(1+\lfloor log_2(k)\rfloor)}\right)$. This opens the avenue for a quantum attack against the consensus mechanism, potentially leading to a quantum 51\% attack against the network. 

Since the signature scheme for ZCash is reliant on the hardness of the discrete logarithm problem, it is susceptible to quantum attacks using Shor's Algorithm. Transactions broadcast to the network could be stolen by a quantum attacker, before they are added to the blockchain. 

The global public parameter that is used in the production of zk-SNARKs, is a public-key that has no corresponding private-key and is reliant on the hardness of the discrete logarithm problem. Quantum attackers through the use of Shor's algorithm could solve to find the global private key. This would not be reliant on the need for any further information as to how the global public parameter was created during the ceremony. With the possession of the global private key, a quantum attacker could create an infinite amount of ZCash tokens. With the private key they would not be able to access other users' transactions being broadcast on the network. However, being able to create tokens at will, especially in a network that has obfuscated transactions this would be extremely dangerous to the network and its associated economy. 

ZCash has a high vulnerability to quantum attacks. Previous examples that we have discussed are at threat to transactions being stolen once broadcast to the network. However, on ZCash the vulnerability allows a quantum assailant to create tokens. Further to this, the work shown by Grassi \textit{et al.}, means the consensus mechanism used by ZCash is also more vulnerable than other blockchain technologies discussed. Transactions that have been broadcast to the network are equally vulnerable, since the signing mechanism relies on the hardness of the discrete logarithm problem.

\section{Conclusion}

This paper demonstrates that the cryptographic schemes that underpin blockchain technologies are highly susceptible to quantum attack, particularly from subgroup-finding-algorithms, such as Shor's algorithm. Table \ref{table} summarizes the vulnerabilities of the blockchain protocols that have been analyzed in this paper, as well as some additional technologies of note that have had in depth analysis. 

Based on the information presented here, we can derive a comparative analysis and ranking of the discussed blockchains.

Of the blockchains analyzed here, we can conclude Monero to be the most secure. This is due to the obfuscation of transacted values on the blockchain. While this obfuscation technique is vulnerable to quantum attacks, as it relies on the difficulty of the discrete logarithm problem, the obfuscation of many transactions may have to be removed in order for the attack to be effective---each time by running Shor's algorithm. Otherwise, a quantum attacker would be blindly attacking transactions that could be of very little financial value. Performing this against many private transactions would be more computationally expensive than similar attacks on other networks. This is further explained in section \ref{sec:monero} on page \pageref{sec:monero}.

The next most secure blockchains are Bitcoin and those with a similar cryptographic structure, like LiteCoin. Transactions that are declared to the network are highly vulnerable to quantum attack from Shor's algorithm. However, under best-use practices, only individual transactions would be under threat, as the reuse of public keys is discouraged in these blockchains. 

In Ethereum, public-key reuse is more common because of its account-based structure. Once a public-key for an account becomes known to the network, that account will be vulnerable to quantum attack. Using a quantum computer the private-key for an account can be computed from the public-key. At this point the attacker can take over the account completely, including the ability to siphon all current funds to another account.

Of the blockchains discussed here, the most vulnerable to quantum attack is ZCash. Despite being a privacy-based blockchain, ZCash uses a public ceremony in order to enable the anonymity of transactions. This creates a global public key based on elliptic-curve cryptography. Using Shor's algorithm, a quantum assailant can easily find this private key. This would allow them to create an unlimited supply of tokens. Furthermore, because the values are obfuscated, this attack would remain unknown to the rest of the network.

Finally, it is worth highlighting the particular issue of Grover algorithm-based attacks. Typically, attacks based on Grover's algorithm are considered less of a threat than those based on Shor's algorithm \citep[see \emph{e.g.}][]{Aggarwal}. However, there is one way in which Grover's algorithm represents a much more serious threat to blockchain technologies than Shor's algorithm does. 

Consider that in order to face the threat of Shor's algorithm one has the option to \emph{swap out} vulnerable cryptography (\emph{e.g.} RSA) to quantum-safe or \emph{post-quantum} cryptography. In the case of PoW there is no such option to swap out the underlying computational problem (from, say, hashing) to one that is not vulnerable Grover speed-up. By definition, PoW requires a computational problem that can be efficiently verified. This necessitates a computational problem that is in NP, which in turn implies the problem is amenable to Grover algorithm speed-up. In short, there is \emph{no possible} PoW system that is not susceptible to Grover speed-up. This implies that quantum actors will always have an advantage over classical ones in PoW-based blockchains---and can use this advantage either to mine more effectively, or as a basis for a $51\%$ attack. The only counter to this being dropping PoW completely in favor of an entirely different system such as Proof of Stake (PoS).

Finally, in conclusion, all blockchain technologies analyzed here have varying, yet ultimately critical vulnerabilities that can be exploited using a sufficiently-developed quantum computer. Fortunately, blockchain technologies are still a fledgling technology, and quantum computers even more so. This gives the industry time to adapt, and course-correct. PoW and many other consensus mechanisms available are sufficiently resistant in the near to medium term from all currently-known quantum algorithms. However, signature schemes for transaction broadcast will need to be changed to use appropriately designed (post-quantum) cryptography in order for a blockchain network to become quantum-safe.

\section*{Acknowledgements} 

The authors of this paper would like to acknowledge the contribution to the early work for this paper from Jeathra Sivarajasingam. The authors would also like to thank Dr. Pauline Bernat and Joanna I. Ziembicka for useful comments during the preparation on this manuscript. CP-D would like to acknowledge funding through the EPSRC Quantum Communications Hub (EP/T001011/1).

\raggedright \Urlmuskip=0mu plus 5mu\relax

\bibliography{article} 
\bibliographystyle{ieeetr}

\end{document}